\documentstyle[axodraw,epsf,12pt,cite]{article}
\newcommand{\epem}{e^+e^-}
\newcommand{\M}{{\cal M}}
\newcommand{\hp}{{\hat p}} \newcommand{\hs}{{\hat s}}
\newcommand{\s}{\sigma}
\newcommand{\beq}{\begin{eqnarray}} \newcommand{\beqa}{\begin{eqnarray*}}
\newcommand{\eeq}{\end{eqnarray}} \newcommand{\eeqa}{\end{eqnarray*}}
\newcommand{\e}{\epsilon} \newcommand{\g}{\gamma}
\newcommand{\bv}{\bar v(k_2)}
\renewcommand{\u}{u(k_1)}
\begin{document}
\begin{center}
{\Large\bf Inclusive vector charmonium production in two-photon $\epem$-annihilation at $\sqrt s=10.6\,\rm GeV$}\\
\vskip 1cm
Luchinsky A. V.\\
\vskip 1cm
{\it Institute for High Energy Phycics, Protvino, 142280 Russia}
\end{center}
\hskip 1cm
\begin{quote}
{\small In this paper we study inclusive production of vector charmonium states $J/\psi$ or $\psi(2S)$ in two-photon $\epem$-annihilation at $\sqrt s=10.6\,\rm GeV$ using color-singlet model. Analytic expressions  for differential cross sections, numerical values of total cross sections and plots of different distributions are presented. These distributions differ significantly from the one-photon case so these modes can be easily separated.
}
\end{quote}

\section{Introduction}

The charmonium states (such as $J/\psi(1S)$, $\psi(2S)$ or $\eta_c$) are of great interest both from theoretical and experimental points of view because of their clear experimental signature and great simplification caused by their non-relativistic nature. In the framework of non-relativistic quantum chromodynamics (NRQCD) \cite{NRQCD,Braaten,Liu} it is possible to describe this processes in a systematic way expressing all quantities as the series in electromagnetic and QCD constants $\alpha$ and $\alpha_s$ and the velocity of the $c$-quark in charmonium meson. In the case of the the the vector charmonia production in $\epem$ annihilation there is also an almost model independent way to describe this process since all non-perturbative constants can be determined phenomenologically from the $(c\bar c)\to\epem$ decay rate.

Among the examples of such reactions is the exclusive double charmonium production in the $\epem$-annihilation, that was subjected to thorough theoretical and experimental investigation. Recently the Belle collaboration has observed the $\epem$ annihilation in two charmonium states at the center-of-mass energy $\sqrt s=10.6\,{\rm GeV}$ \cite{Belle} and their measurements differ significantly from the results obtained in NRQCD \cite{Braaten,Liu,Likh,Qiao,Qiao2,Cho,Kiselev,Baek}. Some efforts were made to explain this discrepancy by using the color-octet mechanism \cite{Chen,Cho2,Qiao2,Baek} or the two-photon contribution to this process \cite{Braaten,Me,Qiao3}, but the results results of these efforts cannot completely explain the difference between theoretical predictions and experimental data.

Another interesting example is the inclusive vector charmonium production in the $\epem$ annihilation. This process was was studied in details, but the contribution from the two-photon intermediate state to this process has not been considered yet. In spite of `the strong suppression caused by the additional factor $\alpha$ it produces completely different distributions over the scattering angle and the energy of charmonium mesons. The subject of this article is to consider this process in almost model-independent way.

In the next section we give the expressions of the coupling constants used in this article and their connection to experimental data. In section 3 analytic expressions for differential cross sections are presented and in the last section we give some numerical results.

\section{Vertices}

The matrix element of the decay of vector meson $V$ into the electron-positron pair can be written in the form
\beqa
\M_V&=&g_V\e_\mu \bv\g^\mu\u,
\eeqa
where $k_1$ and $k_2$ are the electron and positron momenta (their masses are neglected) and $g_V$ is the effective coupling constant, that can be determined from the experiment. The corresponding decay width
\beq
\Gamma^{ee}_V&=&(2\pi)^4\frac{1}{6M_V}\int d\Phi_2(p;k_1,k_2)|\M_V|^2,\label{gee}
\eeq
where $p=k_1+k_2$ is the momentum of the vector meson, $M_V=\sqrt{p^2}$ is its mass and
\beq
d\Phi_n(P;p_1,\dots,p_n)&=&
\delta(P-\sum_{i=1}^n p_i)\prod_{i=1}^n\frac{d^3{\bf p}_i}{(2\pi)^32E_i}\label{LIPS}
\eeq
is $n$-particle Lorentz-invariant phase space. From equation (\ref{gee}) we get
\beq
g_V^2&=&12\pi\frac{\Gamma^{ee}_V}{M_V}.\label{g}
\eeq

The matrix elements of the one-photon annihilation of the electron-positron pair into specific hadronic state $H$ has the form
\beqa
\M(\epem\to\g^*\to H)&=& E_\mu(H)\bv\g^\mu\u=E_\mu(H)L^\mu,
\eeqa
where $E_\mu(H)$ is the effective polarization vector of $H$ which can depend on a number of handrons $n$, their momenta $\hp_i$, spins and masses. The inclusive cross section
\beqa
\s_H&=&\sum_H\s(\epem\to\g^*\to H)=
(2\pi)^4\frac{1}{16(k_1k_2)}L^\mu (L^\nu)^\dag\sum_H
\int d\Phi_n(\hp;\{\hp_i\})E_\mu(H)E^*_\nu(H),
\eeqa
where $\hp=\sum_{i=1}^n\hp_i$ is the total momentum of the hadronic state and $\hs=\hp^2$. The last term in the above equation can depend only on the metric tensor $g_{\mu\nu}$, $\hs$ and $\hp$, but due to Ward identities $L_\mu\hp^\mu=0$ the term proportional to $\hp_\mu\hp_\nu$ cancels and we can rewrite it in the form
\beqa
\sum_H\int d\Phi_n(\hp;\{\hp_i\})E_\mu(H)E_\nu^*(H)=-a(\hs)g_{\mu\nu}.
\eeqa
The function $a(\hs)$ can be splitted in the resonance and the non-resonance parts $a(\hs)=a^r(\hs)+a^{nr}(\hs)$, where
\beqa
a^r(\hs)&=&\sum_V\delta(\hs-M_V^2)g_V^2,
\eeqa
corresponds to the production of a single vector meson $V$ with mass $M_V$, coupling constant $g_V$ is defined according to (\ref{g}) and $\delta(\hs-M_V^2)$ stands for narrow distribution of $\hs$ around $\hs=M_V^2$ with the width~$\sim(\Gamma_V^{ee})^2$,
\beqa
a^{nr}(\hs)&=&\s(\epem\to\mu^+\mu^-)R(\hs)=\frac{\alpha^2}{6\pi^3}\frac{R(\hs)}{\hs}
\eeqa
corresponds to inclusive production of any continuum hadronic state. For the ratio $R(\hs)$ we use the expression
\beq
R(\hs)&=&3\sum_q e_q^2\label{R}
\eeq
where summation is performed over all active quark flavors.

\section{Inclusive production of vector meson}

\begin{figure}
\begin{picture}(400,100)(0,0)
\ArrowLine(10,90)(50,90)\ArrowLine(50,90)(50,10)\ArrowLine(50,10)(10,10)
\Text(5,90)[r]{$e^-$}\Text(5,10)[r]{$e^+ $}
\Text(30,94)[b]{$k_1$}\Text(30,6)[t]{$k_2$}\Text(47,50)[r]{$q_a$}
\Vertex(50,90){1}\Vertex(50,10){1}
\Photon(50,90)(100,90){2}{ 2}\Photon(50,10)(100,10){-2}{2}
\Text(70,95)[b]{$p$}\Text(70,5)[t]{$\hp$}
\Vertex(100,89){3}\Line(102,92)(140,92)\Line(102,87)(140,87)\Vertex(140,89){3}
\Text(145,89)[l]{$V,\e_\mu$}
\Vertex(100,11){3}\Line(100,13)(140,20)\Line(100,8)(140,0)
\Text(142,20)[l]{$\hp_1$}\Text(142,0)[l]{$\hp_n$}\Text(120,10)[l]{$\cdots$}
\Text(155,10)[l]{$H,E_\nu(H)$}

\ArrowLine(260,90)(310,90)\ArrowLine(310,90)(310,10)\ArrowLine(310,10)(260,10)
\Text(255,90)[r]{$e^-$}\Text(255,10)[r]{$e^+ $}
\Text(280,94)[b]{$k_1$}\Text(280,6)[t]{$k_2$}\Text(297,50)[r]{$q_b$}
\Vertex(310,90){1}\Vertex(310,10){1}
\Photon(310,90)(350,10){2}{ 2}\Photon(310,10)(350,90){-2}{2}
\Text(337,75)[b]{$p$}\Text(337,25)[t]{$\hp$}
\Vertex(350,89){3}\Line(352,92)(390,92)\Line(352,87)(390,87)\Vertex(390,89){3}
\Text(395,89)[l]{$V,\e_\mu$}
\Vertex(350,11){3}\Line(350,13)(390,20)\Line(350,8)(390,0)
\Text(392,20)[l]{$\hp_1$}\Text(392,0)[l]{$\hp_n$}\Text(370,10)[l]{$\cdots$}\Text(400,10)[l]{$H,E_\nu(H)$}

\end{picture}
\caption{}\label{fig1} 
\end{figure}

The diagrams of the two-photon $\epem$ annihilation into vector meson $V$ and a specific hadronic state $H$ are shown on figure \ref{fig1}. The corresponding matrix element has the form
\beqa
\M_{VH}=g_V\e_\mu E_\nu(H)\bv\left[\frac{1}{q_a^2}\g^\nu\hat q_a\g^\mu
+\frac{1}{q_b^2}\g^\mu\hat q_b\g^\nu\right]\u=g_V\e_\mu E_\nu(H)L^{\mu\nu}
\eeqa
and the inclusive cross section is
\beq
\s_{VH}&=&\sum_H\s(\epem\to2\g^*\to VH)=\nonumber\\
&=&\frac{2\pi^4g_V^2}{s}\e_\mu\e^*_\alpha\sum_H\int d\Phi_{n+1}(k_1+k_2;p,\{\hp_i\})E_\nu(H)E^*_\beta(H)
L^{\mu\nu}(L^{\alpha\beta})^\dag.\label{svh}
\eeq
The Lorentz-invariant phase space (\ref{LIPS}) satisfies the recursive relation
\beqa
d\Phi_{n+1}(k_1+k_2;p,\{\hp_i\})&=&(2\pi)^3d\hs d\Phi_2(k_1+k_2;p,\hp)d\Phi_n(\hp;\{\hp_i\})
\eeqa
and the expression (\ref{svh}) takes the form
\beq
\s_{VH}&=&\frac{\pi^2g_V^2}{8s}\int\limits_{\hs_{min}}^{\hs_{max}}d\hs\frac{|\bf p|}{\sqrt\hs}a(\hs)
\int\limits_{-1}^1dx|L_{\mu\nu}|^2=
\frac{\pi^2g_V^2}{s}\int\limits_{\hs_{min}}^{\hs_{max}}d\hs\lambda
\left[\frac{\alpha^2}{6\pi^3}\frac{R(\hs)}{\hs}+\sum_{V'}\delta(\hs-M_{V'}^2)g_{V'}^2\right]\times
\nonumber\\&\times&\int\limits_{-1}^1dx
\frac{e_0^2(4-2e_0-e_1e_2)-(e_0^2-6e_0+e_1e_2+4)\lambda^2x^2-\lambda^4x^4}{(e_0^2-\lambda^2x^2)^2}=
\s_{VH}^r+\s_{VH}^{nr},\label{svh2}
\eeq
where $x=\cos\theta$ is the cosine of the scattering angle,
\beqa
\lambda&=&\frac{1}{2}\sqrt{1-\left(\frac{M_V}{\sqrt s}+\frac{\sqrt\hs}{\sqrt s}\right)^2}
\sqrt{1-\left(\frac{M_V}{\sqrt s}-\frac{\sqrt\hs}{\sqrt s}\right)^2},\\
e_0&=&1-\frac{M_V^2}{s}-\frac{\hs}{s},\qquad
e_1=1+\frac{M_V^2}{s}-\frac{\hs}{s},\qquad
e_2=1-\frac{M_V^2}{s}+\frac{\hs}{s},
\eeqa
$\hs_{max}=(\sqrt s-M)^2$ and for $\hs_{min}$ we use two possible values: $\hs_{min}=(2m_\pi)^2$ and $\hs_{min}=1\,{\rm GeV}^2$. In the first case the ratio (\ref{R}) should be corrected by the pion form factor, but its values are close enough to 1 and these corrections are neglected. The term with $V'=V$ in the equation (\ref{svh2}) should be deleted by 2 to avoid double counting.

After integration over the scattering angle we obtain the distribution of non-resonance inclusive cross section over $\hs$:
\beqa
\frac{d\s^{nr}_{VH}}{d\hs}&=&\frac{g^2\alpha^2}{3\pi s\hs}R(\hs)\lambda\left[
\frac{2e_0(1-e_0)-e_1e_2+\lambda^2}{e_0^2-\lambda^2}+
\frac{e_0^2-2e_0+2}{e_0\lambda}\log\frac{e_0+\lambda}{e_0-\lambda}\right].
\eeqa

The integration over $\hs$ was performed numerically and its results are presented in the next section.

\section{Numerical results}

\begin{figure}
\begin{picture}(400,270)
\put(10,10){\epsfxsize=14cm \epsfbox{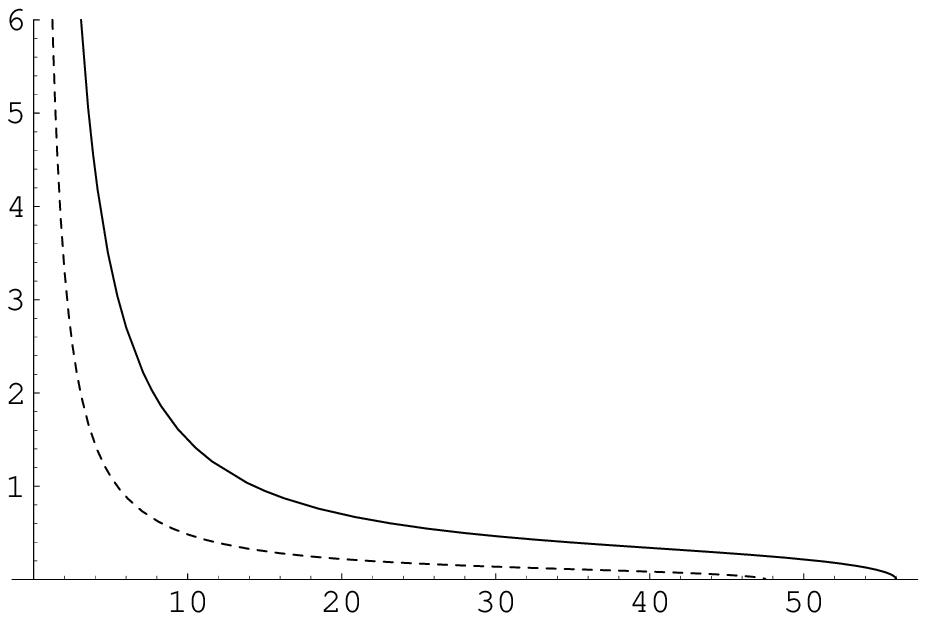}}
\put(30,260){$\frac{d\sigma^{nr}(\epem\to V H)}{d\hs}\,,{\rm \frac{fb}{GeV^2}}$}
\put(380,10){$\hs\,,{\rm GeV^2}$}
\end{picture}
\caption{Distribution over the invariant mass of the hadronic state for $V=J/\psi$ (solid line) and $V=\psi(2S)$ (dashed line)}
\label{fig2}
\end{figure}

\begin{figure}[p]
\begin{picture}(400,270)
\put(10,10){\epsfxsize=14cm \epsfbox{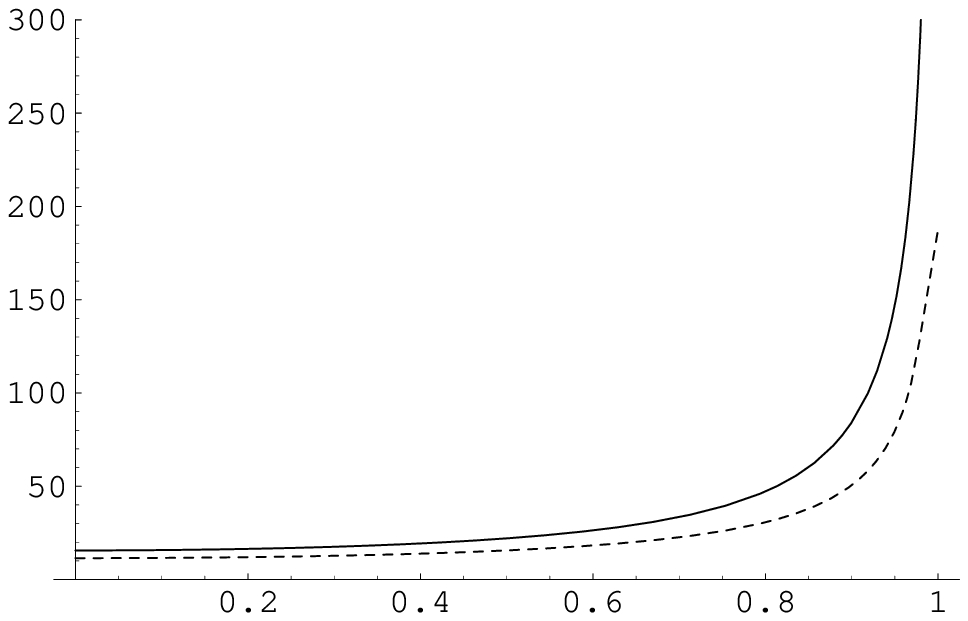}}
\put(30,260){$\frac{d\sigma^{nr}(\epem\to J/\psi H)}{dx}\,,{\rm fb}$}
\put(400,10){$x$}
\end{picture}
\caption{Angular distribution at $\hs_{min}=(2m_\pi)^2$ (solid line) and  $\hs_{min}=1\,{\rm GeV}^2$ (dashed line)}
\label{fig3}
\end{figure}

\begin{figure}[p]
\begin{picture}(400,270)
\put(10,10){\epsfxsize=14cm \epsfbox{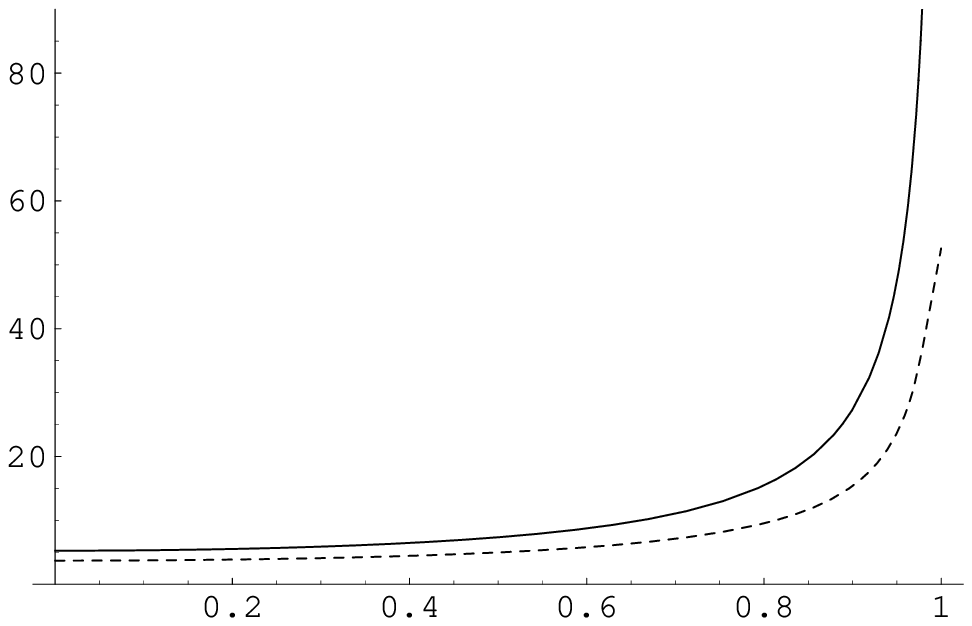}}
\put(30,260){$\frac{d\sigma^{nr}(\epem\to \psi(2S) H)}{dx}\,,{\rm fb}$}
\put(400,10){$x$}
\end{picture}
\caption{Angular distribution at $\hs_{min}=(2m_\pi)^2$ (solid line) and  $\hs_{min}=1\,{\rm GeV}^2$ (dashed line)}
\label{fig4}
\end{figure}

On figure \ref{fig2} we show distributions of inclusive non-resonance cross sections $\s_{VH}^{nr}$ over the invariant mass of the hadronic state for $V=J/\psi$ and $\psi(2S)$. As one can easily see, it is peaked near the small values of $\hs$ (or equivalently in the end region of the vector meson energy in the center-of-mass frame). This form of distribution is explained by the virtual photon propagators and differ significantly from the case of inclusive vector charmonium production in the one-photon $\epem$ annihilation. On figures \ref{fig3} and \ref{fig4} angular distributions of inclusive $J/\psi$ and $\psi(2s)$ for two different values of $\hs_{min}$ are shown. Again we observe the strong peak near $x=\pm1$ in contrast to the one-photon case. Finally in table 1 the values of total resonance and non-resonance cross sections of $J/\psi$ and $\psi(2S)$ inclusive production for different values of $\hs_{min}$ are presented. They are smaller then the one-photon induced cross sections by about order of magnitude, but the specific form of distributions could help to separate this signal.

Author would like to thank A.K.Likhoded for useful discussions. This work was supported by Russian Foundation of Basic Research, grants 33-02-16558 and 00-15-96645.

\begin{table}[t]
$$\begin{array}{|c|c|c|c|}
\hline
V & \s^r_{VH}\, ,{\rm fb} & \s^{nr}_{VH}\, {\rm fb}, \hs_{min}=(2m_\pi)^2&
\s^{nr}_{VH}\, {\rm fb}, \hs_{min}=1\,{\rm GeV}^2 \\
\hline
J/\psi & 58.0 & 99.7  & 56.0 \\
\psi(2S) & 20.0 & 32.0 & 16.1 \\
\hline
\end{array}$$
\label{tab}\caption{}
\end{table}

\end{document}